# Evolving Tsukamoto Neuro-Fuzzy Model for Multiclass Covid-19 Classification with Chest X-Ray Images

Marziyeh Rezaei[1], Sevda Molani[*,2], Negar Firoozeh[3], Hossein Abbasi[4], Farzan Vahedifard[5], Maysam Orouskhani[3]

[1] Research Assistant, Department of Electrical and Computer Engineering, University of Washington
Marziyeh@uw.edu

[2] Postdoc Research, Institute of Systems Biology (Corresponding author)
Smolani@systemsbiology.org

[3] Postdoctoral Research Fellow, Department of Radiology, University of Washington
{Maysam, Nfirooze@uw.edu}

[4] South Tehran Branch, Islamic Azad University
Hossein.abbasi48@gmail.com

[5] Department of Diagnostic Radiology and Nuclear Medicine, Rush Medical College
Farzan_vahedifard@rush.edu

**Abstract**

Due to rapid population growth and the need to use artificial intelligence to make quick decisions, developing a machine learning-based disease detection model and abnormality identification system has greatly improved the level of medical diagnosis. Since COVID-19 has become one of the most severe diseases in the world, developing an automatic COVID-19 detection framework helps medical doctors in the diagnostic process of disease and provides correct and fast results. In this paper, we propose a machine learning-based framework for the detection of Covid-19. The proposed model employs a Tsukamoto Neuro-Fuzzy Inference network to identify and distinguish Covid-19 disease from normal and pneumonia cases. While the traditional training methods tune the parameters of the neuro-fuzzy model by gradient-based algorithms and recursive least square method, we use an evolutionary-based optimization, Cat swarm algorithm to update the parameters. In addition, six texture features extracted from chest X-ray images are given as input to the model. Finally, the proposed model is conducted on the chest X-ray dataset to detect Covid-19. The simulation results indicate that the proposed model achieves an accuracy of 98.51%, sensitivity of 98.35%, specificity of 98.08%, and F1- score of 98.17%.

## 1. Introduction

The use of machine learning algorithms in medicine has become increasingly popular in recent years. It is possible to detect abnormalities more quickly through machine learning algorithms in medical imaging by analyzing the features of the input images. In this way, doctors can use medical imaging to diagnose any disease using machine learning-based frameworks. In the past three years, COVID-19 has become the most common infection in the world and reducing COVID-19-related deaths will require the development of a fast, stable, and accurate detection method.

Both neural networks and fuzzy inference systems are powerful machine learning methods with applications in system identification and prediction.

The most applicable intelligent networks are fuzzy-neural systems with both the abilities of a neural network and a fuzzy inference system [1]. Fuzzy reasoning is developed to implement fuzzy implications relations. Four commonly used fuzzy reasoning methods are:
- Mamdani-type fuzzy reasoning.
- Larsen-type fuzzy reasoning.
- TSK-type fuzzy reasoning.
- Tsukamoto-type fuzzy reasoning.

Among the proposed neuro-fuzzy systems, Tsukamoto-type neural fuzzy inference networks, TNFIN, and adaptive neuro-fuzzy inference systems, ANFIS, are the most widely used models in many engineering fields, including drought and rainfall forecasting, agriculture, and solar energy modeling [2,3,4]. Researchers have implemented neuro-fuzzy systems as powerful machine learning models like the deep neural networks developed recently in medicine, including disease detection [5,6]. For example, [7] has created an ANFIS-based classifier for brain tumor characterization, and [8] proposed a model for breast cancer detection by using the ANFIS classifier.

However, it is imperative to train the parameters of neuro-fuzzy models to improve their accuracy. The antecedent and consequent parameters of ANFIS and TNFIN are both nonlinear. Therefore, finding an optimal learning step size and calculating gradients in each step is difficult, which causes the convergence process to be slow. The Tsukamoto neuro-fuzzy inference system, TNFIN, is a five-layer feed-forward neural fuzzy network that uses fuzzy reasoning of the Tsukamoto type. To improve accuracy, TNFINs are trained using the LSE algorithm and the GD method. After each iteration, the error function is monitored versus the step size to determine the optimal learning step size. The antecedent and consequent parameters of TNFIN are both nonlinear. Therefore, it isn't easy to decide on an optimal learning step size and calculate gradients in each step, which causes the convergence process to be slow. Evolutionary computation algorithms have gained popularity in recent years for finding better solutions based on population analysis with different stochastic techniques. As a result, evolutionary algorithms can be implemented to solve the training problem. Evolutionary algorithms can find better solutions based on population than traditional training methods, using different stochastic techniques. In this paper, we use the Cat Swarm Optimization (CSO) [9] algorithm to train the parameters of TNFIN. We selected the CSO due to its high accuracy, exploration, and exploitation ability compared to state-of-the-art algorithms.

In the past few years, many researchers have attempted to implement machine learning algorithms into an automated system to assist the radiologist and medicine in diagnosing COVID-19 [10-16]. Although computed tomography and magnetic resonance imaging are standard methods of diagnosing [17], chest X-rays offer a quicker and cheaper method of detecting COVID-19. These medical images provide incomplete and uncertain information, making this data interpretation difficult. Fuzzy systems can easily interpret these types of information due to the ability of fuzzy models to deal with unclear information. This is due to the realistic presentation of imprecise details in images, expressing this information at various stages with fuzzy sets, and the manifestation of heterogeneous components [18]. In this paper, we propose a TNFIN model based on CSO to detect Covid-19 using features provided by X-ray images. The model is given six types of components extracted from the X-ray dataset. Then, the model's weights are updated by CSO.

The rest of the article is organized: Section 2 reviews machine learning algorithms for COVID-19 detection. In addition, the cat swarm algorithm and its evolutionary process are explained. Section 3 introduces the methods, including TNFIN and feature extraction. Finally, section 4 presents the simulations and experimental results of running the proposed model for detecting COVID-19.

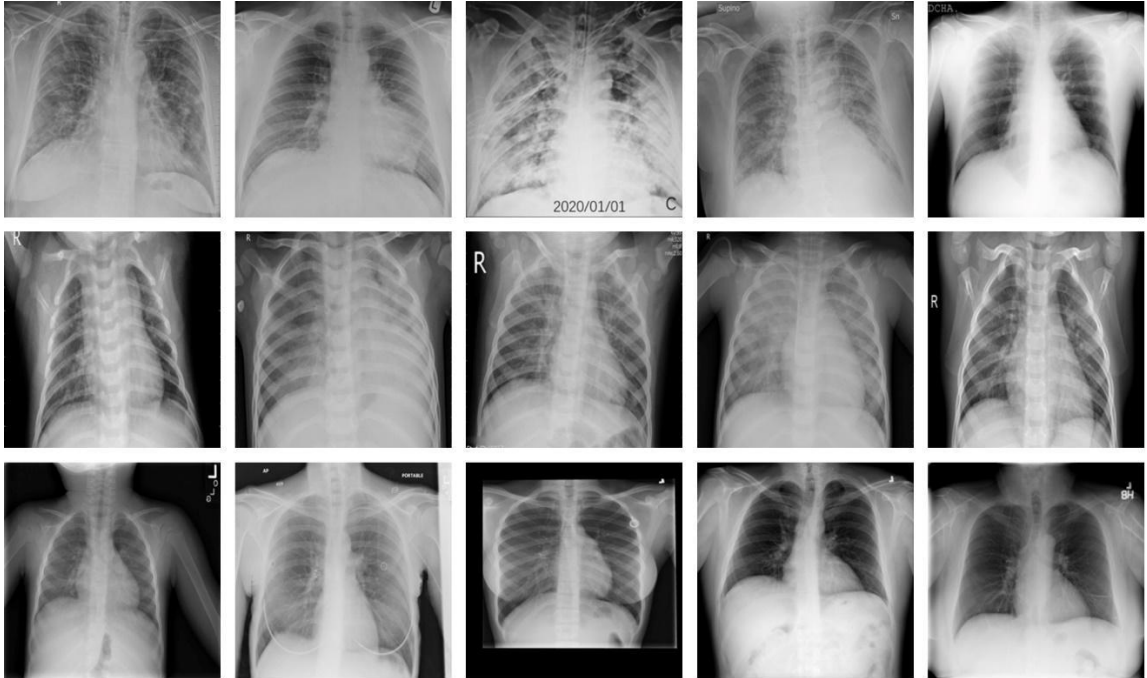

Fig 1. First row: images of Covid-19 cases, Second row: images of Pneumonia cases, Third row: normal cases [12]

## 2. Background

### 2.1. Covid-19 Detection using Machine Learning Algorithms

Some researchers have developed techniques to diagnose COVID-19 in a timely manner and prevent its severe consequences since the beginning of the pandemic [10, 11]. [12] proposed a model using a convolution neural network (CNN) and a pre-trained AlexNet algorithm for COVID-19 detection. The models were tested on 170 chest X-rays and 361 CT images of COVID-19 cases. The accuracy of CNN and pre-trained AlexNet was 94% and 98% for X-ray images and 94.1% and 82% for CT images, respectively. Although the model's accuracy is reasonable, the tested dataset was small. [13] developed a convolutional neural network model for automatic COVID-19 detection. The random forest, support vector machine, decision tree, and AdaBoost classifiers were combined to achieve better accuracy. The model was evaluated using 660 positive COVID-19 chest X-ray images with 98.91% accuracy. Another model for the detection of COVID-19 is proposed in [14].

In this model, features are extracted using a Convolutional Neural Network and classifiers are constructed using Support Vector Machines. We tested the proposed model on 306 X-rays (102 COVID-19s and 204 normal and pneumonia) to classify them into three categories (standard, COVID-19, and pneumonia). In separating normal from COVID-19 and pneumonia from COVID-19, the method achieved an accuracy of 97.33%. [15] developed a COVID-19 detection algorithm that relies on CNN for deep feature extraction and long short-term memory (LSTM) for classification. They used a large dataset with 4575 X-ray images (1525 COVID-19 cases) for training and evaluation. An accuracy of 99.4% was achieved, which outperforms other works considering the large dataset examined. [16] used ANFIS architecture based on the subtractive clustering technique for automatic COVID-19 detection from chest X-ray images, benefiting from operating with small datasets while maintaining high accuracy. A simpler architecture than CNN-based models enabled the model to achieve 98.7% accuracy when analyzing 2000 chest X-rays.

## 2.2. Cat Swarm Optimization

In order to find global solutions, a variety of algorithms are used. There are some cases in which optimization algorithms are developed using swarm intelligence. The development of many algorithms that mimic the swarm behavior of creatures has been widespread in recent years, including Ant Colony Optimization, which mimics ant behavior, Particle Swarm Optimization, which mimics bird behavior, and Cat Swarm Optimization, which mimics cat behavior.

In the field of swarm intelligence, the Cat Swarm Optimization (CSO) algorithm is a novel algorithm. Researchers have used this algorithm in a wide variety of applications. [9]. In the CSO algorithm, the cat's behavior is categorized into two different kinds of modes: 'Seeking mode 'and 'Tracking mode'. In CSO, every cat has its own position which consists of d dimensions, velocities for each dimension, fitness values, which represent the cat's accommodation to the fitness function, and a flag to indicate its seeking or tracking mode. Ultimately, the most efficient solution would be to place one of the cats in the right position. Throughout the iterations, the CSO keeps the preferred solution. The Cat Swarm Optimization algorithm has two modes to solve the following problems:

### 2.2.1. *Seeking Mode*

The algorithm uses the seeking mode to model the cat's behavior during periods of rest or alertness. It is time to think and decide what the next step should be.
Further, because of its structure, the algorithm emphasizes global search ability, and the main objective is to find multiple solutions. This mode can be described as an exploration phase, which explores the search space on a global scale. There are four main parameters in this mode: seeking a memory pool (SMP), seeking a range of selected dimensions (SRD), counting dimensions to change (CDC), and self-position consideration (SPC). Described as follows, the seeking mode involves [24]:

**Step 1:** Make j copies of $cat_k$'s current position, where j = SMP. Let j = (SMP-1) if SPC is true, then keep the present position as a candidate.

**Step 2:** Randomly add or subtract SRD percent from the present values for each copy and replace the old ones.

**Step 3:** Calculate fitness values (FS) for all candidate points.

**Step 4:** Calculate the selecting probability of each candidate point based on Equation (1) if all FS are not exactly equal, otherwise set all the selecting probabilities of each candidate point to 1.

**Step 5:** Pick a random point from the candidate points and replace the position of $cat_k$. If the fitness function aims to find the minimum solution, $FS_b = FS_{max}$, otherwise $FS_b = FS_{min}$.

$$P_i = \frac{|FS_i - FS_{max}|}{FS_{max} - FS_{min}}$$
(1)

### 2.2.2. *Tracing Mode*

The second mode of the algorithm is the tracing mode. This mode is used by cats to track their food and targets. The algorithm exploits solutions through local search, which is called exploitation. The tracing mode procedure is described as follows in Section [24]:

**Step 1:** Update velocity values for every dimension using Equation (2).

**Step 2:** Check if the velocity is within the maximum range. Whenever a new velocity exceeds the range, it is set to the limit.

**Step 3:** Adjust the cat$_k$'s position according to Equation (2)

(2)
$$V_{k,d} = V_{k,d} + r_1 c_1 (X_{best,d} - X_{k,d})$$
$$X_{k,d} = X_{k,d} + V_{k,d}$$

X$_{best,d}$ is the position of the cat with the best fitness value, X$_{k,d}$ is the position of cat$_k$, c$_1$ is the acceleration coefficient for extending the cat's velocity and is usually 2.05, and r$_1$ is a random value uniformly generated between 0 and 1. Equation (3) represents adaptive inertia weight and new velocity update equations where 'w$_s$' is the first inertia weight, 'i' is the current iteration, and 'i$_{max}$' is the maximum iteration.

(3)
$$W(i) = W_s + \frac{(i_{max} - i)}{2 \times i_{max}}$$
$$V_{k,d} = W(i) \times V_{k,d} + r_1 c_1 \times (X_{best,d} - X_{k,d})$$

## 3. Methods

In this section, we propose the Evolving-TNFIN model for Covid-19 detection. First, the standard TNFIN model is introduced. Then, the training phase indicates how all nonlinear parameters of TNFIN are trained using CSO. Finally, six texture features derived from chest X-ray images are calculated.

### 3.1. Tsukamoto-Type Neural Fuzzy Inference Network (TNFIN)

Both neural networks and fuzzy logic are model-free estimators and share similar abilities to deal with uncertainties and noise [25]. In addition, both encode information in parallel and distributed architectures within a numerical framework. Hence, it is possible to convert a fuzzy logic architecture into a neural network and vice versa. This makes it possible to combine the advantages of neural networks and fuzzy logic [25]. A network built in this way could use the powerful training algorithms that neural networks have at their disposal. This allows them to obtain the required parameters not available in the fuzzy logic architecture. Moreover, the network obtained in this way would not remain a black box because it would have fuzzy logic interpretation capabilities regarding linguistic variables [26]. TNFIN combines two approaches, neural networks, and fuzzy systems. The Tsukamoto-type neural fuzzy network has five layers with the same type of nodes. Its architecture is shown in Figure 2. A simple two-input one-output system is chosen to emphasize and explain the basic ideas of the hybrid learning algorithm [27].

Five layers of TNFIN are explained as follows: Layer1 is the fuzzification layer. This layer executes a fuzzification process and makes fuzzy outputs. Fuzzification process will be done by membership functions (MFs) as follows:

(4)
$$\mu A_i(x) \frac{1}{1 + \left(\frac{x - c_i}{a_i}\right)^2}$$

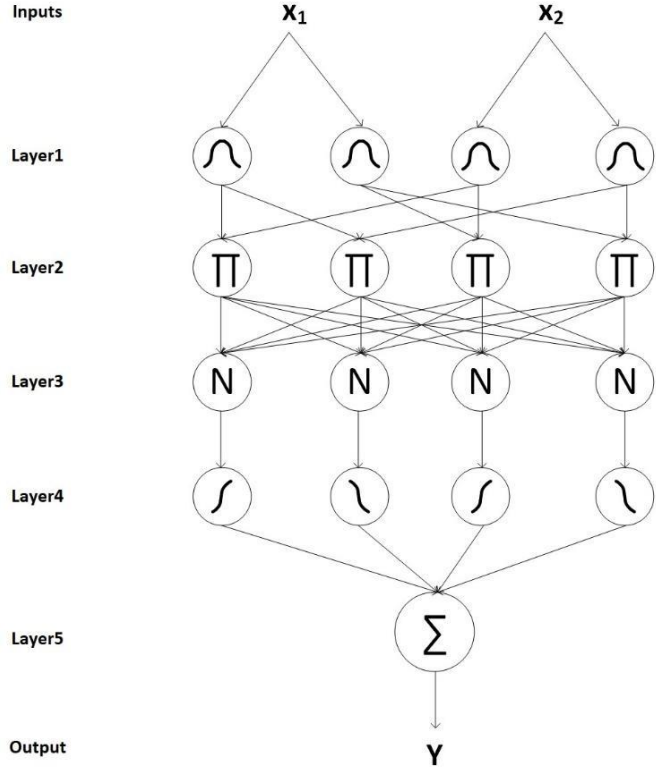

Fig 2. Tsukamoto-type neural fuzzy inference system with 2 inputs and 1 output

Here, $c_i$ and $a_i$ are known as primary nonlinear parameters. The second layer executes the fuzzy AND of the antecedent part of the fuzzy rules and the output of this layer multiplies the input signals. Each neuron in this layer calculates the related firing strength.

(5)
$$O_{2,i} = w_i = \mu A_i(x) \times \mu B_i(y), i = 1,2$$

The third layer normalizes the membership functions and neurons in this layer calculate the related normalized firing strength which shows the effect of a rule in result.

(6)
$$O_{3,i} = \overline{w}_i = \frac{w_i}{w_1 + w_2}, i = 1,2$$

The fourth layer executes the conclusion part of the fuzzy rules. This layer uses Tsukamoto fuzzy reasoning for the defuzzification process. Figure (3) shows the Tsukamoto fuzzy defuzzification. In this layer, m and s are called consequent parameters and are used to adjust the shape of the membership function of the consequent part. The output of this layer is:

(7)
$$O_k^4 = O_k^3 y_k = \begin{cases} O_k^3 \left( c_k - d_k \sqrt{\frac{1}{O_k^3} - 1} \right) & \text{If } k = odd \\ O_k^3 \left( c_k + d_k \sqrt{\frac{1}{O_k^3} - 1} \right) & \text{If } k = even \end{cases}$$

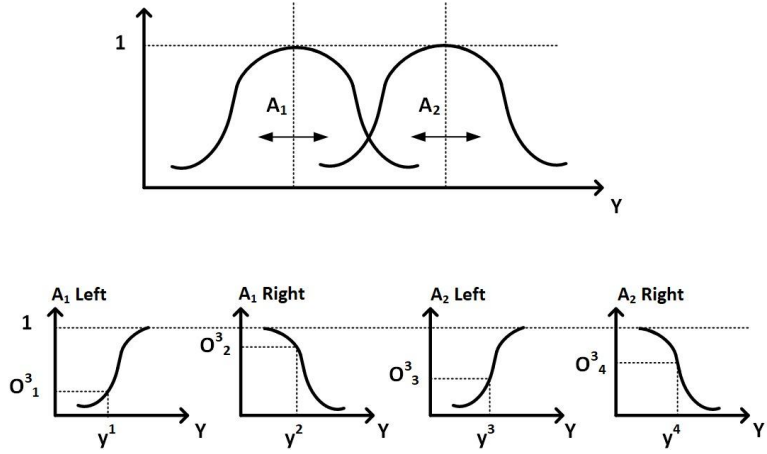

Fig 3. Tsukamoto-type defuzzification

$$Y = \frac{O_1^3 \cdot y_1 + O_2^3 \cdot y_2 + O_3^3 \cdot y_3 + O_4^3 \cdot y_4}{O_1^3 + O_2^3 + O_3^3 + O_4^3}$$

Layer 5 computes the output of the fuzzy system by summing up the outputs of the fourth layer which is the defuzzification process.

(8)
$$O^5 = \sum O_k^4$$

The general learning procedure in neural fuzzy systems consists of modifying their parameters by presenting them with the input data and desired output data. Typically, by adjusting the parameters of fuzzy membership functions and the weights of connections between different layers, a certain performance index is optimized. Here the goal is to minimize the following error function.

(9)
$$E = \frac{1}{2} \sum_{p=1}^{n} \left(T(p) - O^5(p)\right)^2 = \sum_{p=1}^{n} E_p$$

Where $O^5(p)$ is the network output for the $P^{th}$ training point, $T(p)$ is the corresponding targeting output, and n is the total number of the points in the training data set.

The TNFIN training process is a hybrid training method including the Least Square method (LSE) and Gradient Descend (GD). In the training phase, each iteration of running the hybrid training algorithm is composed of a forward pass and a backward pass. The consequent parameters are identified in the forward pass by means of the LSE method and the antecedent parameters are tuned in the backward pass using the GD method. However, the nonlinearity and complexity of the gradient calculation causes slow convergence. As an example, 'm' as one of the last layer's parameters is trained as follows:

(10)
$$m_{t+1} = m_t - \eta \frac{\partial E}{\partial p_k} = p_k - \eta \times \left(\frac{\partial E}{\partial o_5} \times \frac{\partial o_5}{\partial o_4} \times \frac{\partial o_4}{\partial p}\right)$$

### 3.2. Training Phase

This section introduces a new training method for the training of TNFIN. While the standard TNFIN's parameters, including antecedent and consequent parameters, are updated, and trained through the gradient-based optimization algorithms, the nonlinearity of the parameters may cause the model to be stuck in a local optimum. Unsimilar to gradient-based optimizers that need the gradient information, evolutionary-based algorithms try to find the global solutions of a problem via an evolutionary process. In this research, the

Cat swarm optimization algorithm is employed to tune the nonlinear parameters of TNFIN. In the proposed model, the antecedent parameters in layer two and consequent parameters of layer four are trained by CSO. The main goal of using CSO is to minimize the loss of the model during the training process of TNFIN. Here, 'y' is the target output and '$y_{out}$' is the estimated output.

$$Error = \frac{1}{2}(y - y_{out})^2 \qquad (11)$$

In the simulations of this paper, TNFIN uses three membership functions for each input and the relation of finding the rules number is described in Equation (13). Table 1 describes the number of parameters trained by CSO. Also, the CSO parameters are shown in Table 2. As an example, we can train the consequent parameter 'm' using CSO as follows: It is notable that we consider an extra parameter as a velocity parameter ($m_v$) for each TNFIN's parameter that can be adjusted by CSO.

$$m_v(t) = m_v(t-1) + r_1 c_1 (m_{best} - m(t-1)) \qquad (12)$$
$$m(t) = m(t-1) + m_v(t)$$

Table 1. Training parameters in proposed learning algorithm (* **Total number of parameters: 753**)

| Type | Parameter | Parameter Number |
|---|---|---|
| Antecedent | c | $NeuronsNumber \times InputDimension = 18$ |
| Antecedent | a | $InputDimension = 6$ |
| Consequent | m,s | $Rules\ number = (NeuronsNumber)^{InputDimension} = 729$ |

Table 2. Parameters of CSO

| Parameter | Range or Value |
|---|---|
| SMP | 3 |
| SRD | 10% |
| CDC | 100% |
| Mixture Ratio | 50% |
| r1 | [0,1] |
| Initial Inertia Weight | 0.15 |
| Iteration | 200 |
| Population Size | 40 |
| Epoch in each iteration | 5 |

*3.3. Feature Extraction*

As the first step of pre-processing, each image in the data set is converted to gray scale and resized to 512 × 512 then they are normalized to decrease computational complexity. We used a gray-level co-occurrence matrix (GLCM) technique to extract the features [15]. The gray-level co-occurrence matrix (GLCM) is a statistical method of examining texture based on the spatial relationship between pixels. The step of this algorithm is as follows: the first step of the GLCM algorithm is to quantize the image data by calculating the GLCM matrix from the test image. This step initializes a two-pixel GLCM matrix and generates a normalized symmetric matrix. Using this version, the correlation (the pixel correlation over the whole image), contrast (the intensity contrast between each pixel and its neighbor over the full image), energy (the total amount of the GLCM square parameters) and homogeneity (the distributed similarity of the elements in the GLCM to its diagonal) features are calculated [16]. In addition, we computed mean and variance (the average intensity value and the way the pixels are spread) from the histogram distribution of each image. The mathematical formulas for six texture features are listed below. The rows and columns of the GLCM matrix are shown by i,j. In addition, p (i,j) denotes the relative position of the pixels in the matrix. The average and standard deviation of each image's pixels are indicated by δ and μ.

$$Contrast = \sum_{i=1}^{n}\sum_{j-=1}^{n}(i,j)^2 \times p(i,j), \quad Correlation = \sum_{i,j=1}^{n} p_{i,j}\frac{(i-\mu)(j-\mu)}{\delta^2}$$

$$Energy = \sum_{i=1}^{n}\sum_{j-=1}^{n} p(i,j)^2, \quad Homogeneity = \sum_{i,j=1}^{n}\frac{p_{ij}}{1+(i-j)^2}$$

$$Mean = \sum_{i=1}^{n} i \times p(i), \quad Standard\ Deviation = \sqrt{\sum_{i=1}^{n}\sum_{j-=1}^{n}(i,j)^2 \times p(i,j)}$$

(13)

## 4. Experiments

In this section, we conduct the proposed evolving TNFIN model on the X-ray covid dataset. Dataset description, evaluation metrics, and simulation results are explained as follows.

*4.1. Experimental Setup*

In this study we employed chest x-ray dataset from different sources [14]. The chest X-ray images of 341 COVID-19 patients have been obtained from the open-source GitHub repository shared by Dr. Joseph Cohen et al. [28]. This repository is consisting of chest X-ray/computed tomography (CT) images of mainly patients with acute respiratory distress syndrome (ARDS), COVID-19, Middle East respiratory syndrome (MERS), pneumonia, severe acute respiratory syndrome (SARS). 2800 normal (healthy) chest X-ray images were selected from "ChestX-ray8" database [29]. In addition, 1493 viral pneumonia chest X-ray images were used from Kaggle repository called "Chest X-Ray Images (Pneumonia)" [30]. We resized the images to ones with a resolution of 224 × 224 pixels.

To do the comparison between TNFIN and Evolving-TNFIN we used all images as a three-class problem while to have a fair comparison between our proposed model and the state-of-the-art methods we divided the dataset to two classes including the images affected by Covid-19 and the non-affected images. During the training process, we selected 1000 samples while 80 percent was considered as training and remaining as test data. In addition, all experiments consisted of 10 independent cycles, and each cycle consisted of 500 iterations. Also, the simulation was coded and implemented by MATLAB software and in a PC that has an Intel Processor of 2.7 GHz, Quad core CPU, and 12 GB of Memory.

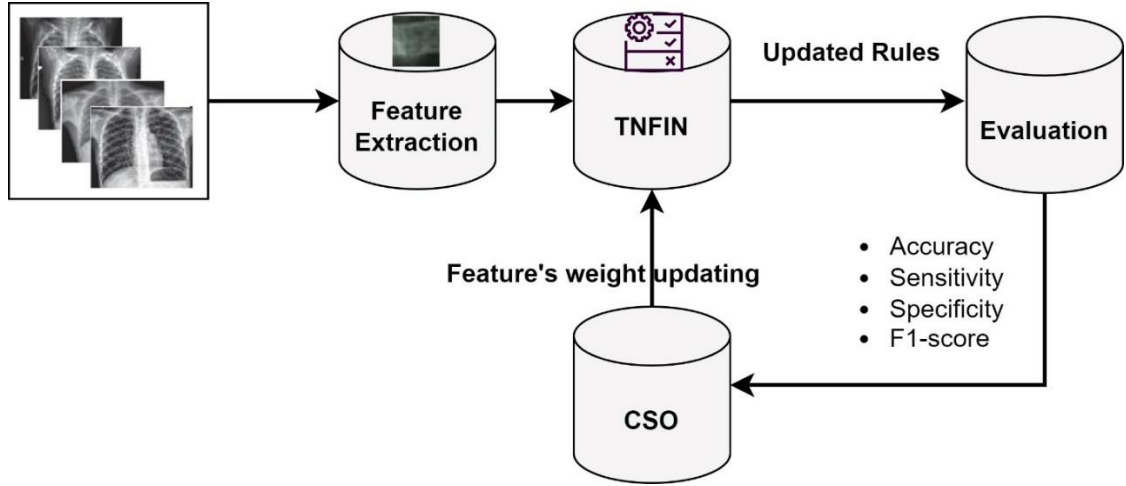

Fig 4. Automated process of Covid-19 classification. Feature extraction extracts the texture features of input. While TNFIN generates the corresponding rules, its parameters are updated through CSO. In addition, four performance metrics evaluate the performance of TNFIN.

*4.2. Evaluation Metrics*

We employ four measures including Accuracy, Sensitivity, Specificity, and F1-score to measure the performance of the proposed model and compare the obtained results provided by the model with state-of-the-art methods. TP and FP indicate the correct and incorrect prediction of the model when the case is positive (Covid-19). As for the negative cases (normal or pneumonia), the correct prediction is denoted by TN while FN indicates the incorrect prediction of the model.

(14)

$$Accuracy = \frac{TP + TN}{TN + FP + TP + FN} \qquad Sensitivity = \frac{TP}{TP + FN}$$

$$Specificity = \frac{TN}{TN + FP} \qquad F1 - score = \frac{2 \times TP}{2 \times TP + FP + FN}$$

*4.3. Results*

In this section, we conduct classification and disease detection with X-Ray images using TNFIN with three architectures: original TNFIN, evolving TNFIN with constant inertia weight, and adaptive weight. There are three classes including Normal, Pneumonia, and Covid-19 patients. The comparison of proposed model with other state-of-the-art models are summarized in Table 3.

As the results show, evolving TNFIN achieved higher performance than standard TNFIN in terms of Accuracy, Sensitivity, Specificity, and F-1 score. It is notable that both models achieved the highest performance and did the best classification for the Normal category. Furthermore, figure 5 illustrates the training-test mean squared error and error histogram with 20 Bins. Table 4 indicates the comparison between the results obtained by the proposed model and seven methods that have been already compared by [15]. Since the results provided by [15] is for a two-class problem, we conducted our model for a binary classification problem including 1) infected by Covid and 2) not infected by Covid. The simulation results denotes that our proposed model outperforms the following methods as well: ANFIS + GLCM, modified AlexNet + Shallow CNN + Resnet50, and DarkNet. Among all introduced methods in Table 3, EMCNet achieved the best accuracy.

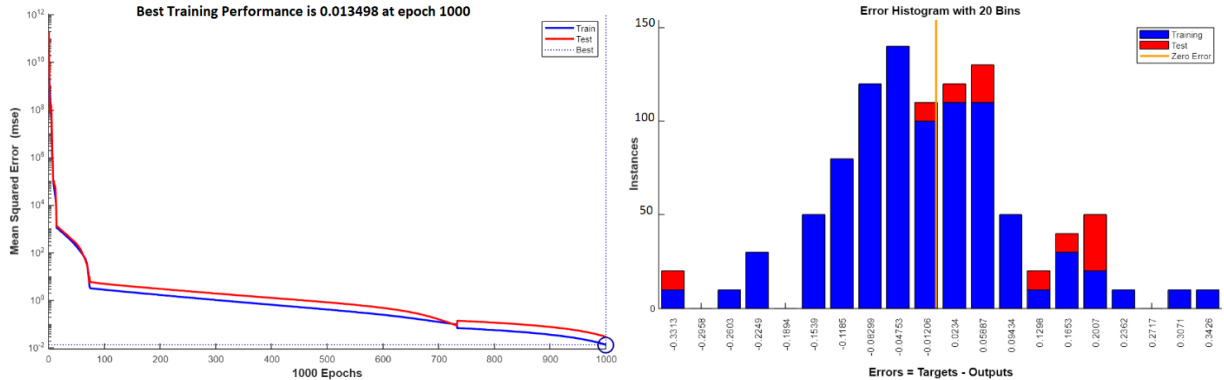

Fig 5. Up: Training and test mean squared error: The mean squared error is the average squared difference between the target output and predicted output, down: Error Histogram with 20 Bins

Table 3. Results comparison between the proposed model and state-of-the-art algorithms. The results for a binary problem include two classes

| Method | Accuracy |
| --- | --- |
| Darknet | 98.08% |
| COVIDX-Net | 60–90% |
| ResNet50, InceptionV3 and Inception- ResNetV2 | 98%, 97%, 87% |
| Shallow CNN | 96.92% |
| Modified AlexNet, simple CNN | 94.40% |
| EMCNet | 98.91% |
| ANFIS + GLCM | 98.67% |
| Evolving TNFIN with constant weight | 98.70 |
| Evolving TNFIN with adaptive weight (Our method) | 98.81 |

*4.4. Ablation Study*

In this section, we analyze the effect of each contribution in the performance of the final model. The original TNFIN model uses the traditional gradient descend training algorithm to update the weights. While using CSO with constant inertia weight to update the TNFIN's parameters improves its performance, adaptive inertia weight aims to increase the global and local search ability of CSO simultaneously. As a result, we provide an inertia weight with large value at first iterations to explore the feasible solutions and try to decrease it gradually to increase the exploitation ability of CSO and give more attention to the local solutions. Table 4 shows the summary of results for four performance metrics achieved by the original TNFIN, evolving TNFIN with constant and adaptive weight. The results indicate that evolving TNFIN with adaptive inertia weight outperforms two the models.

Table 4. Mean and standard deviation (mean ± std) of results achieved by TNFIN, Evolving-TNFIN with constant weight and adaptive weight.

| Class | Metric | TNFIN | Evolving-TNFIN Constant w=0.9 | Evolving-TNFIN adaptive w |
|---|---|---|---|---|
| Normal | Accuracy | 98.89 ± 0.00001 | 99.01 ± 0.00001 | **99.30 ± 0.000001** |
| | Sensitivity | 97.81 ± 0.00017 | 99.54 ± 0.00019 | **99.68 ± 0.00036** |
| | Specificity | 96.25 ± 0.033 | **99.56 ± 0.018** | 99.39 ± 0.010 |
| | F1 | 97.17 ± 0.008 | 99.01 ± 0.0007 | **99.64 ± 0.00061** |
| Pneumonia | Accuracy | 95.27 ± 0.0024 | 96.21 ± 0.0039 | **98.63 ± 0.0034** |
| | Sensitivity | 96.17 ± 0.0066 | 97.25 ± 0.0081 | **98.47 ± 0.0080** |
| | Specificity | 94.27 ± 0.018 | **98.27 ± 0.0069** | 98.24 ± 0.0055 |
| | F1 | 96.01 ± 0.008 | 97.67 ± 0.003 | **98.33 ± 0.0079** |
| Covid-19 | Accuracy | 95.89 ± 0.001 | 96.32 ± 0.0031 | **98.51 ± 0.0021** |
| | Sensitivity | 96.45 ± 0.0062 | 97.18 ± 0.0066 | **98.35 ± 0.0071** |
| | Specificity | 94.5 ± 0.027 | 97.19 ± 0.0064 | **98.08 ± 0.0051** |
| | F1 | 96.62 ± 0.009 | **98.23 ± 0.003** | 98.17 ± 0.0057 |

Table 5. p-values of Kruskal-Wallis test for the results of table 4

| | Accuracy (%) | Sensitivity (%) | Specificity (%) | F-1 (%) |
|---|---|---|---|---|
| Covid-19 | **2.32E-4** | **5.1E-3** | **8.27E-4** | **9.94E-4** |
| Pneumonia | **2.61E-4** | **4.81E-3** | **6.18E-4** | **5.76E-3** |
| Normal | 0.062 | **0.047** | **2.6E-4** | **3.37E-3** |

Table 6. Results of Mann-Whitney U test

| Metric | Model | TNFIN | Evolving-TNFIN with constant w | Evolving-TNFIN with adaptive w |
|---|---|---|---|---|
| Accuracy | TNFIN | n/a | | |
| | Evolving-TNFIN with constant w | 0.042 | n/a | |
| | Evolving-TNFIN with adaptive w | 9.1E-4 | 1.9E-3 | n/a |
| Sensitivity | TNFIN | n/a | | |
| | Evolving-TNFIN with constant w | 0.039 | n/a | |
| | Evolving-TNFIN with adaptive w | 0.017 | 1.5E-3 | n/a |
| Specificity | TNFIN | n/a | | |
| | Evolving-TNFIN with constant w | 3.8E-3 | n/a | |
| | Evolving-TNFIN with adaptive w | 1.9E-4 | - | n/a |
| F-1 | TNFIN | n/a | | |
| | Evolving-TNFIN with constant w | 3.1E-3 | n/a | |
| | Evolving-TNFIN with adaptive w | 2.9E-3 | - | n/a |

*4.5. Statistical Analysis*

This section uses the Kruskal-Wallis test to consider the statistical analysis that was done on the values of four performance metrics including Accuracy, Sensitivity, Specificity, and F-1. A significant value indicates at least one sample stochastically dominates one other sample. Therefore, this test was performed to determine whether there was a statistically significant difference (confidence level of 95%) between the values obtained by three models. The obtained p-values by Kruskal-Wallis test are presented by Table 5, where bold values indicate the statistically significant between the results achieved by three models. Table 5 shows that there is a statistically significant difference between the results obtained from three models except for the accuracy of normal class. As a result, the evolving TNFIN significantly outperforms the original TNFIN in different metrics. In addition to the Kruskal-Wallis test, Mann-Whitney U test was performed to determine which model statistically provides significant difference. This test is also performed under a confidence level of 95%. Tables 6 presents the results of the Mann-Whitney U tests in each dataset.

Here, "-" indicates no statistically significant difference was found. Table 6 shows there is almost a significant difference between the results provided by the three models in dataset. As an exception, there is no significant difference for the Specificity and F-1 achieved by "Evolving-TNFIN with constant weight" and "Evolving-TNFIN with adaptive weight".

**Conclusions**

In this study an evolutionary-based Tsukamoto neuro-fuzzy framework was proposed to detect Covid-19. In contrast to the standard training algorithm, Cat swarm algorithm was employed to train the nonlinear parameters of the model. In addition, texture features from chest X-ray images were extracted and given to the model. The proposed model was conducted on the chest X-ray dataset to detect the Covid-19 disease. The dataset was treated as a three-class problem including normal, Pneumonia, and Covid-19 cases. The experiments illustrated that our proposed model achieved 98.81% for Accuracy metric and outperformed most state-of-the-art methods except EMCNet.